\documentclass[prd,twocolumn,superscriptaddress,amssymb,amsfonts,amsmath,nofootinbib,preprintnumbers]{revtex4-2}
\usepackage{graphicx,slashed,xcolor,multirow,bbold,mathtools,sidecap,tikz,bm,ulem,enumitem,booktabs,array,amsmath,amssymb}
\usepackage[colorlinks=true,linktoc=page,linkcolor=purple,citecolor=teal,urlcolor=magenta]{hyperref}
\usepackage[compat=1.1.0]{tikz-feynman}
\usepackage{siunitx,adjustbox,comment}
\definecolor{lime}{HTML}{A6CE39}
\DeclareRobustCommand{\orcidicon}{\hspace{-2.1mm}
\begin{tikzpicture}
\draw[lime,fill=lime] (0,0.0) circle [radius=0.13] node[white] {{\fontfamily{qag}\selectfont \tiny ID}}; \draw[white,fill=white] (-0.0525,0.095) circle [radius=0.007]; 
\end{tikzpicture} \hspace{-3.5mm} }
\foreach \x in {A, ..., Z} {\expandafter\xdef\csname orcid\x\endcsname{\noexpand\href{https://orcid.org/\csname orcidauthor\x\endcsname} {\noexpand\orcidicon}}}

\let\emph\textit

\begin{document}

\preprint{TTP25-038, P3H-25-085, ZU-TH 69/25, ICPP-100}

\title{Constraining $A\to ZH$ with $H\to t\bar t$ in the Low-Mass Region}

\author{Saiyad Ashanujjaman\orcidA{}}
\email{saiyad.ashanujjaman@kit.edu}
\affiliation{Institut f\"ur Theoretische Teilchenphysik, Karlsruhe Institute of Technology, Engesserstra\ss e 7, D-76128 Karlsruhe, Germany}
\affiliation{Institut f\"ur Astroteilchenphysik, Karlsruhe Institute of Technology, Hermann-von-Helmholtz-Platz 1, D-76344 Eggenstein-Leopoldshafen, Germany}

\author{Guglielmo Coloretti\orcidE{}}
\email{guglielmo.coloretti@protonmail.com}
\affiliation{Physik-Institut, Universität Zürich, Winterthurerstrasse 190, CH–8057 Zürich, Switzerland}

\author{Andreas Crivellin\orcidB{}}
\email{andreas.crivellin@cern.ch}
\affiliation{Physik-Institut, Universität Zürich, Winterthurerstrasse 190, CH–8057 Zürich, Switzerland}

\author{Siddharth P.~Maharathy\orcidC{}}
\email{siddharth.prasad.maharathy@cern.ch}
\affiliation{School of Physics and Institute for Collider Particle Physics, University of the Witwatersrand, Johannesburg, Wits 2050, South Africa}
\affiliation{Indian Institute of Science Education and Research Pune, Dr.~Homi Bhabha Road, Pune 411008, India}

\author{Bruce Mellado\orcidD{}}
\email{bmellado@mail.cern.ch}
\affiliation{Institute of High Energy Physics, 19B, Yuquan Road, Shijing District, 100049, Beijing, China.}
\affiliation{School of Physics and Institute for Collider Particle Physics, University of the Witwatersrand,
Johannesburg, Wits 2050, South Africa}

\begin{abstract}
The decay $A\to ZH$ is a characteristic signal of two-Higgs-doublet models (2HDMs), where $A$ and $H$ lie primarily within the same $SU(2)_L$ multiplet, leading to a coupling of order $g_2$ to the $Z$ boson. The subsequent decay $H\to tt^{(*)}$ is particularly promising, as it gives rise to distinct final states involving multiple leptons and $b$-jets. The required splitting between $m_A$ and $m_H$ can naturally occur near the electroweak scale while being consistent with perturbative unitarity. Whereas dedicated ATLAS and CMS searches focused on the region with both top-quarks on-shell, we cover lower masses where one top quark is off-shell by recasting Standard Model $t\bar{t}Z$ measurements of ATLAS and CMS. The obtained limits on $\sigma(A\to ZH)\times {\rm Br} (H\to t\bar t)$ are between 0.12\,pb and 0.62\,pb. Interestingly, we observe these stringent limits despite a preference (up to $2.5\sigma$) for a non-zero new physics signal, most pronounced around for $m_A \approx 450$--460\,GeV and $m_H\approx 290$\,GeV, with a best-fit value of $\sigma(A \to ZH) \times {\rm Br}(H \to t\bar t) \approx 0.3$\,pb. This cross section can be accommodated within a top-philic 2HDM for a top-Yukawa coupling of the second Higgs doublet of $0.16 \lesssim \mu_t \lesssim 0.33$.
\end{abstract}

\maketitle

\section{Introduction}
With the observation of the Brout-Englert-Higgs boson~\cite{Higgs:1964ia, Englert:1964et, Higgs:1964pj, Guralnik:1964eu} at the LHC~\cite{Aad:2012tfa, Chatrchyan:2012ufa}, the particle content of the Standard Model (SM) has been confirmed. While the measured properties of this $125$\,GeV scalar boson are consistent with SM predictions~\cite{Langford:2021osp, ATLAS:2021vrm}, the existence of additional Higgs bosons is not excluded, provided their contribution in electroweak symmetry breaking is subdominant and their mixing with the SM Higgs is small. In fact, numerous extensions of the SM Higgs sector have been proposed, including $SU(2)_L$ singlets~\cite{Silveira:1985rk, Pietroni:1992in, McDonald:1993ex}, doublets~\cite{Lee:1973iz, Fayet:1974pd, Fayet:1977yc, Haber:1984rc, Kim:1986ax, Peccei:1977hh, Turok:1990zg}, triplets~\cite{Ross:1975fq, Konetschny:1977bn, Cheng:1980qt, Lazarides:1980nt, Schechter:1980gr, Magg:1980ut, Mohapatra:1980yp, Gunion:1989ci, Rizzo:1990uu, Chardonnet:1993wd, Blank:1997qa}, and even higher representations.

The search for new Higgs bosons is a primary goal of the LHC program. Of particular interest are the cascade decays of heavy Higgs states (see Refs.~\cite{ATLAS:2024itc,CMS:2024phk} for an overview), such as the decay of a pseudoscalar Higgs boson $A$ into a $Z$ boson and a new scalar $H$ ($A\to ZH$).
Notably, this chain decay has been identified as a `smoking gun' signature for a strong first-order electroweak phase transition, which could explain the origin of the matter-antimatter asymmetry in the Universe~\cite{Dorsch:2014qja, Goncalves:2022wbp, Biekotter:2023eil}.
While we focus on the decay $A \to ZH$, a strong first-order electroweak phase transition can also be realized when the channel $H \to ZA$ is kinematically open, even though parameter scans often favor the former~\cite{Basler:2016obg, Dorsch:2017nza}. From a collider perspective, both decay chains lead to very similar final states when followed by $A,H \to t\bar t$.\footnote{The two processes can nevertheless be distinguished using spin correlations~\cite{Arco:2025ydq}.}
Furthermore, owing to their clean experimental signatures, these processes provide powerful probes of extended Higgs sectors. In particular, the decay $H\to t\bar t$ yields distinctive multi-lepton final states with $b$-jets, and has been searched for above the top threshold by ATLAS~\cite{ATLAS:2023zkt} and CMS~\cite{CMS:2024yiy}.\footnote{ATLAS considers $m_H>350$\,GeV and CMS $m_H>330$\,GeV. Since CMS does not mention a $t\bar t$ threshold, it seems that they used the $\overline{\text{MS}}$ mass for the top quark, whereas for the decay width the on-shell mass would be more appropriate. We therefore show the ATLAS search region in our plots.}

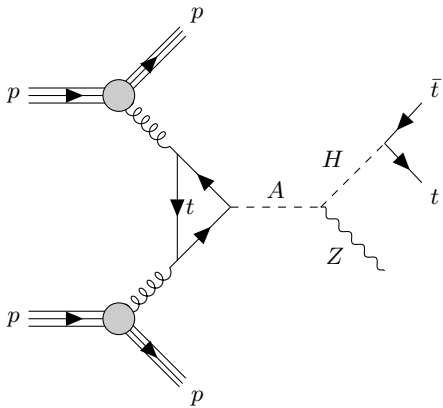
\begin{figure}[t!]
\centering
\begin{tikzpicture}[baseline=(current bounding box.center)]
\begin{feynman}
\vertex (a);
\vertex [above left=2.1cm of a] (c);
\vertex [above=0.1cm of c] (cu); 
\vertex [below=0.1cm of c] (cd);
\vertex [below left=2.1cm of a] (d); 
\vertex [above=0.1cm of d] (du); 
\vertex [below=0.1cm of d] (dd);
\vertex [above left=1.cm of a] (c1);
\vertex [below left=1.cm of a] (d1); 
\vertex [left=1.2cm of c] (p1) {$p$}; 
\vertex [left=1.2cm of cu] (p1u); 
\vertex [left=1.2cm of cd] (p1d);
\vertex [left=1.2cm of d] (p2) {$p$}; 
\vertex [left=1.2cm of du] (p2u); 
\vertex [left=1.2cm of dd] (p2d);
\vertex [above right=1.2cm of c] (pp1) {$p$}; 
\vertex [above right=1.15cm of cu] (pp1u); 
\vertex [above right=1.26cm of cd] (pp1d); 
\vertex [below right=1.2cm of d] (pp2) {$p$}; 
\vertex [below right=1.26cm of du] (pp2u); 
\vertex [below right=1.14cm of dd] (pp2d); 
\vertex [right=1.2cm of a] (b) ;
\vertex [above right=1.2cm of b] (e);
\vertex [below right=1.2cm of b] (f);
\vertex [above right=0.7cm of e] (g) {$\bar t$}; 
\vertex [below right=0.7cm of e] (h) {$t$};
 
\diagram{
(p1) -- [fermion] (c) -- [gluon] (c1), (p1u) -- (cu), (p1d) --  (cd),
(p2) -- [fermion] (d) -- [gluon] (d1),  (p2u) -- (du), (p2d) -- (dd),
(c) -- [fermion] (pp1), (cu) -- (pp1u), (cd) -- (pp1d),
(d) -- [fermion] (pp2), (du) -- (pp2u), (dd) -- (pp2d),
(c1) -- [fermion, edge label=$t$] (d1) -- [fermion] (a)  -- [fermion] (c1),
(a) -- [scalar, edge label=$A$] (b),
(f) -- [boson, edge label=$Z$] (b) -- [scalar, edge label=$H$] (e),
(g) -- [fermion] (e) -- [fermion] (h),
};
\end{feynman}
\node[draw, circle, minimum size=12pt, inner sep=0pt, fill=gray!40] at (c) {};
\node[draw, circle, minimum size=12pt, inner sep=0pt, fill=gray!40] at (d) {};
\end{tikzpicture}
\caption{Feynman diagram depicting the process $pp\to A\to ZH$ with $H\to t\bar t$, leading to a $t\bar{t}Z$-like signature.}
\label{fig:Feynman}
\end{figure}

However, the low-mass region below the top threshold, where one of the top quarks is off-shell, remains essentially unexplored. In this Letter, we close this gap by investigating the process $pp\to A\to ZH$ with $H\to t\bar t$ (see Fig.~\ref{fig:Feynman}), focusing on the mass range where one top quark is on-shell while the other is virtual. For this, we recast the ATLAS and CMS analyses of SM $t\bar tZ$ production~\cite{ATLAS:2023eld, CMS:2024mke}, which provide powerful experimental handles for such signatures~\cite{Ashanujjaman:2025una}. 

\begin{figure}[t!]
\centering
\includegraphics[width=0.46\textwidth]{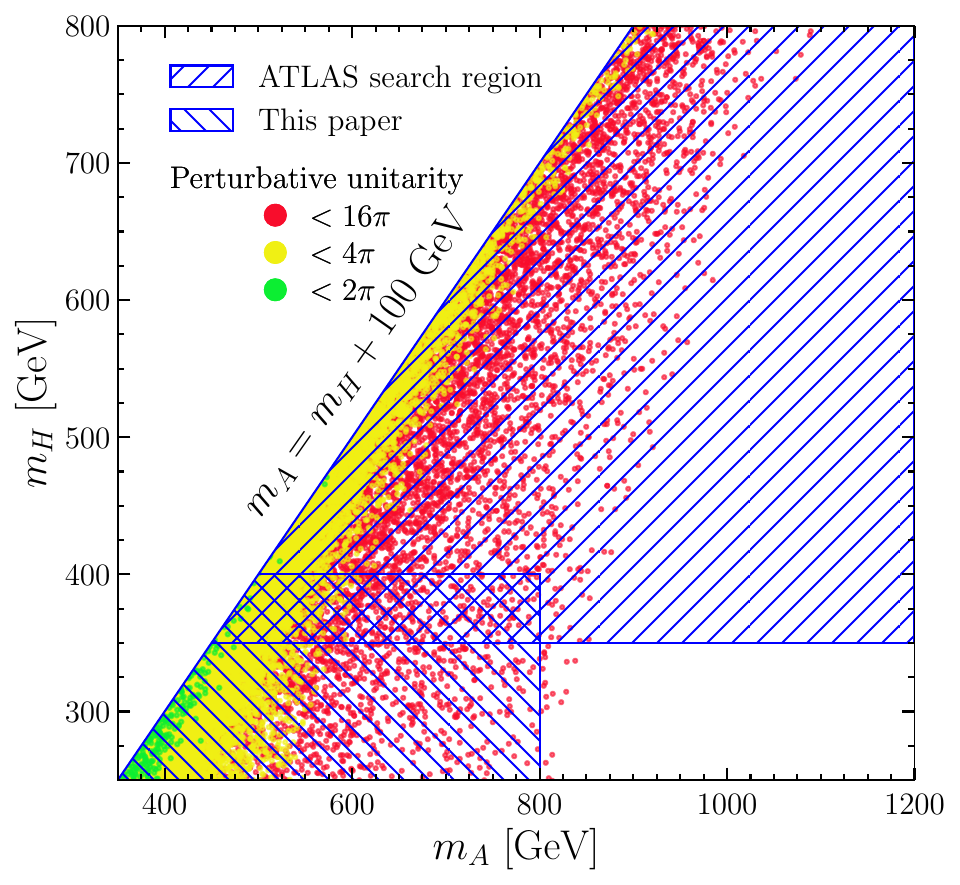}
\caption{Points allowed by EW precision data and perturbative unitarity for different upper limits on the tree-level scattering amplitudes: $<16\pi$ (red), $<4\pi$ (yellow) and $<2\pi$ (green). The search region from ATLAS, as well as the one covered in this work, is hatched. One can see that while only part of the ATLAS region contains viable points, our region is fully populated and closes the gap for $A\to ZH$ with $H\to t\bar t$ searches. }
\label{fig:perturbativity}
\end{figure}

From a theoretical perspective, the mass splitting among the components of an $SU(2)_L$ multiplet is governed by the electroweak (EW) vacuum expectation value (VEV) $v$: $m_A^2=m_H^2+\mathcal{O}(v^2)$. Consequently, for masses near the EW scale a gap exceeding the $Z$ boson mass arises naturally while being consistent with perturbative unitarity, whereas for higher masses, such splittings are increasingly disfavored. This behavior is illustrated in Fig.~\ref{fig:perturbativity}, which shows the regions allowed by EW precision data and perturbative unitarity for different upper limits on the tree-level scattering amplitude. While a sizable part of the ATLAS search region does not satisfy these constraints, the low-mass region explored in this Letter is fully populated by viable points in parameter space.

\section{Analyses of $t\bar{t}Z$ differential distributions}
We consider a $CP$-odd Higgs boson $A$ produced via gluon fusion at the LHC with the subsequent decays $A\to ZH$ and $H\to t\bar t$, where one of the top quarks is off-shell. The resulting $t\bar t Z$-like signature enables us to exploit the measurements of the differential $t\bar tZ$ (and $tWZ$) cross sections by CMS~\cite{CMS:2024mke} and ATLAS~\cite{ATLAS:2023eld} to constrain this beyond-the-SM scenario.

The analysis targets final states with three or four leptons and at least one $b$-tagged jet, corresponding to $t\bar tZ$ (and $tWZ$) production in the SM. The CMS analysis provides results unfolded to the parton level (after radiation but before hadronization) and presents differential cross sections for five kinematic observables: the $Z$ boson transverse momentum ($p_T(Z)$), the transverse momentum of the lepton from the $W$ boson ($p_T(\ell_W)$), the azimuthal angle between the two $Z$ leptons ($\Delta\phi(\ell^+,\ell^-)$), the angular separation between the $Z$ boson and the $W$ boson lepton ($\Delta R(Z,\ell_W)$), and the cosine of the angle between the $Z$ boson and the negatively charged lepton from its decay ($\cos\theta^*_Z$). The ATLAS analysis reports only $t\bar tZ$ differential cross sections, applying a more stringent requirement on the invariant mass of the opposite-sign lepton pair, unfolded to both particle and parton levels, covering 15 observables listed in Table~15 of Ref.~\cite{ATLAS:2023eld}.

To validate our setup, we simulate the SM processes $pp \to t\bar tZ$ and $tWZ$ using {\tt MadGraph5\_aMC\_v3.5.3}~\cite{Alwall:2014hca, Frederix:2018nkq} with the {\tt NNPDF31\_nlo\_as\_0118\_1000} parton distribution function~\cite{NNPDF:2017mvq} at next-to-leading order (NLO) in QCD.\footnote{At NLO, $tWZ$ production interferes with the leading order $ttZ$ process. To take this into account, we use the {\tt MadSTR} plugin~\cite{Frixione:2019fxg}, which removes overlap at the amplitude level using the diagram removal approach.} The obtained parton-level events are interfaced with {\tt Pythia 8.3}~\cite{Sjostrand:2014zea} using the {\tt CMS-CUETP8S1-CTEQ6L1} tune~\cite{CMS:2015wcf} to simulate particle decays, parton showering, and radiation.\footnote{We previously applied the same strategy to constrain the $WZ$ decay mode of a charged Higgs produced from top decays~\cite{Ashanujjaman:2025una}.} 

\begin{figure*}[htb!]
\centering
\includegraphics[width=0.48\textwidth]{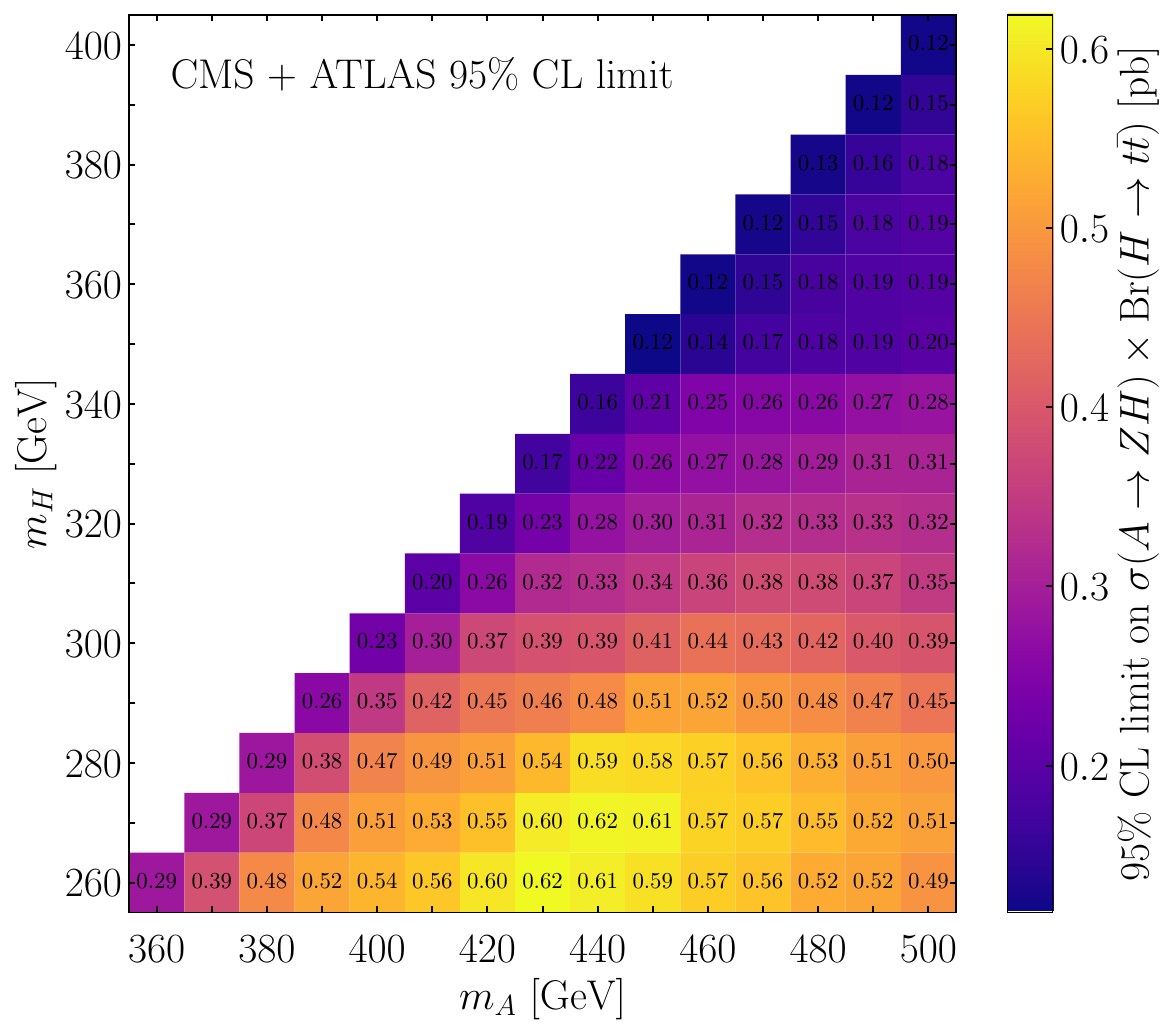} ~~
\includegraphics[width=0.48\textwidth]{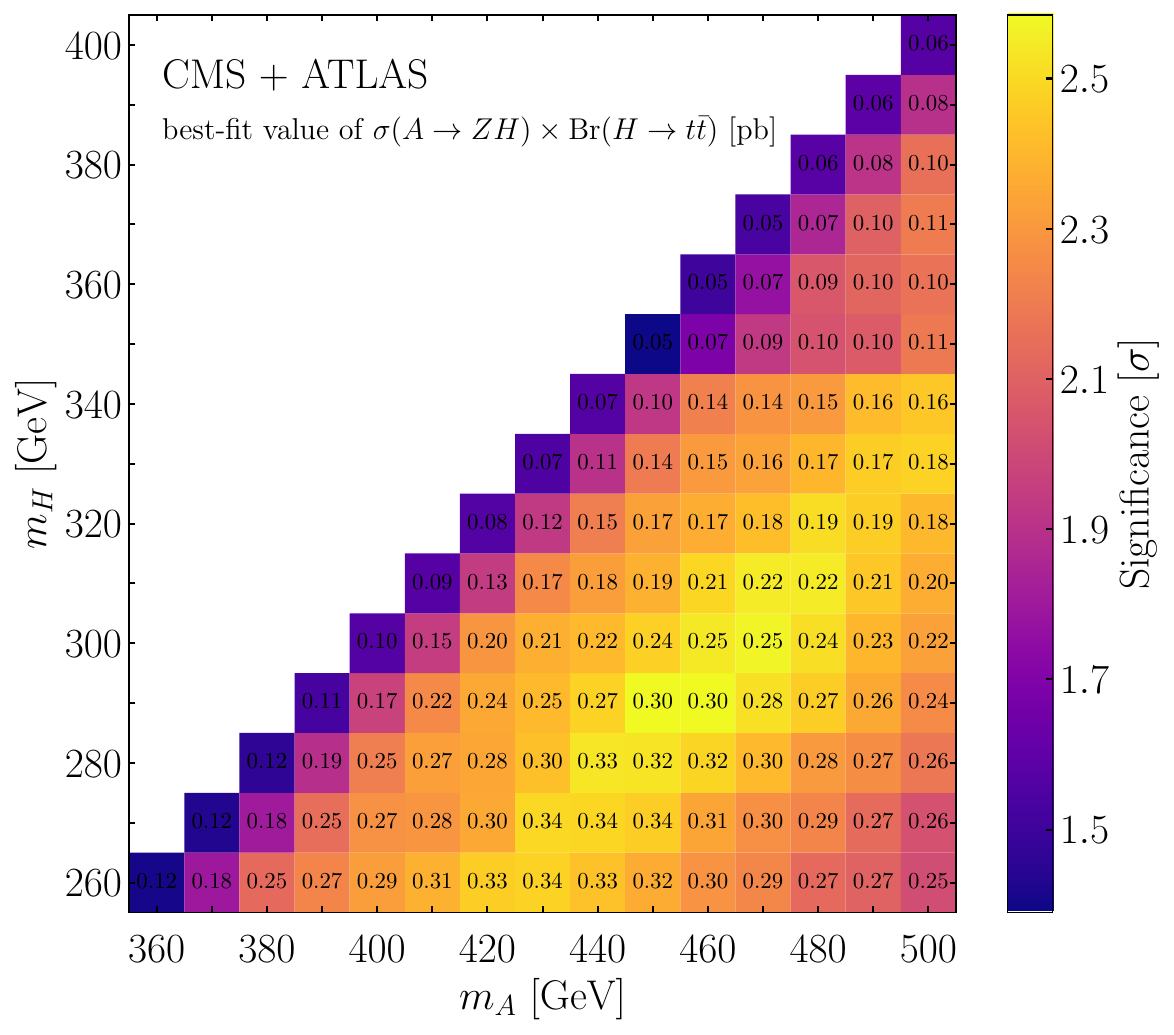}
\caption{95\% CL upper limit (left) and best-fit (right) value of $\sigma(A \to ZH) \times {\rm Br}(H \to t\bar t)$ in units of pb in the $m_A$--$m_H$ plane with $m_A-m_H \geq 100$\,GeV, obtained by combining the CMS and ATLAS analyses. The color bar in the left (right) plot indicates the 95\% CL upper limit for the cross section (the preference for a non-zero NP signal in units of standard deviation). An extended region with $m_A< 800\,\mathrm{GeV}$ is provided in the Supplemental Material.}
\label{fig:limit_combined}
\end{figure*}

For the reconstruction and selection of the leptons (electrons and muons) and jets (including $b$-tagged jets), we closely follow the CMS and ATLAS analyses; jets are clustered with the anti-$k_T$ algorithm~\cite{Cacciari:2008gp} implemented in {\tt FastJet 3.3.4}~\cite{Cacciari:2011ma}. Events are required to contain at least three leptons, including a same-flavor opposite-sign pair with an invariant mass in the nominal $Z$ boson window, and a third lepton consistent with a $W$ boson decay. All additional selection requirements---jet and $b$-tagged jet multiplicities and kinematic thresholds on leptons and jets---are applied to reproduce the event selections used in the CMS and ATLAS signal regions, and are summarized explicitly in the Supplemental Material.

We then simulate our new physics (NP) signal $pp\to A\to ZH$ with $H\to t\bar t$ using the same setup. To have an on-shell $Z$ boson, we require a mass splitting $m_A-m_H \geq 100$\,GeV. Our scan covers $m_H$ from 260\,GeV to 400\,GeV (such that one top quark and the $W$ boson originating from the off-shell top quark are on-shell) and $m_A$ from 360\,GeV to 800\,GeV. The resulting differential distributions for the benchmark point $(m_A,m_H)=(460,290)$\,GeV and $\sigma(A \to ZH) \times {\rm Br}(H \to t\bar t) = 0.3$\,pb are shown in in the Supplemental Material.

\section{Results and Interpretation}
The statistical model for the analysis is built from the binned data templates, SM predictions, and the NP contribution. The NP signal strength is extracted through a simultaneous $\chi^2$ fit:
\begin{align*}
\chi^2 = [\sigma_i^{\rm data} - \sigma_i^{\rm theory} ] \; \Sigma_{ij}^{-1} \; [\sigma_j^{\rm data} - \sigma_j^{\rm theory} ]\,,
\end{align*}
where $i,j$ run over the bins across all observables, $\Sigma_{ij}$ denotes the covariance matrix, $\sigma_i^{\rm data}$ is the measured cross section in bin $i$, and
\begin{align*}
\sigma_i^{\rm theory} = \mu_{\rm SM} \, \sigma_i^{\rm SM} + \mu_{\rm NP} \, \sigma_i^{\rm NP},
\end{align*}
represents the expected cross section in bin $i$, with SM and NP contributions weighted by the corresponding fit parameters; $\mu_{\rm NP}$ is identified with $\sigma(A \to ZH) \times {\rm Br}(H \to t\bar t)$. Interference effects between the SM and NP contributions are neglected; given the narrow width of $A$ in the parameter space considered, these effects are expected to be subleading compared to current experimental uncertainties. Correlations among the differential observables are obtained from our SM simulation, following Ref.~\cite{Banik:2023vxa}.

For the CMS analysis, the theoretical uncertainties are given and incorporated by adding them in quadrature with the experimental ones, allowing us to fix $\mu_{\rm SM} = 1$. In contrast, the ATLAS measurement does not include or provide theoretical uncertainties. We thus use the {\tt MG5\_aMC@NLO + Pythia 8} SM prediction, which lies in between the two simulations obtained from {\tt SHERPA} (without and with multi-leg merging of additional partons) and take into account the theory error by profiling over $\mu_{\rm SM}$ by allowing a 5\% deviation from 1, which corresponds approximately to the known uncertainty of the inclusive $t\bar t$ production cross section~\cite{Czakon:2011xx}.

Using the global $\chi^2$ (the sum of ATLAS and CMS), we extract model-independent limits on $\sigma(A \to ZH) \times {\rm Br}(H \to t\bar t)$ in the $m_A$--$m_H$ plane at 95\% confidence level (CL). The resulting exclusion is shown in the left panel of Fig.~\ref{fig:limit_combined}. Furthermore, as illustrated in the right panel, the fit exhibits a mild (up to 2.5$\sigma$) preference for a non-zero NP signal, most pronounced around for $m_A \approx 450$--460\,GeV and $m_H\approx 290$\,GeV, with a best-fit value of $\sigma(A \to ZH) \times {\rm Br}(H \to t\bar t) \approx 0.3$\,pb. Our limits are in good agreement with the dedicated ATLAS search for $A \to ZH$ with $H \to t\bar{t}$ in the overlap region, which demonstrates that reinterpretations of $t\bar tZ$ data can provide competitive sensitivity to heavy-Higgs signatures. For instance, at $m_A = 450(500)$\,GeV and $m_H = 350(400)$\,GeV, ATLAS reports a 95\%~CL limit of $\sigma(A \to ZH) \times \mathrm{Br}(H \to t\bar{t}) \approx 0.13(0.14)~\mathrm{pb}$ in the narrow-width approximation{\footnote{In the ATLAS interpretation~\cite{ATLAS:2023zkt}, results are provided for $\tan\beta = 0.5, 1,$ and $5$ and restricted to the narrow-width regime $\Gamma_A/m_A \le 25\%$. Since our limits in Fig.~\ref{fig:limit_combined} are derived in the narrow-width approximation without fixing $\tan\beta$, we compare them to the ATLAS results at $\tan\beta = 5$, which lies safely within this regime.}~\cite{ATLAS:2023zkt}, while our combined fit yields $\approx 0.12~\mathrm{pb}$, consistent with the expected improvement from including the CMS data.

\begin{figure*}[htb!]
\centering
\includegraphics[width=0.48\textwidth]{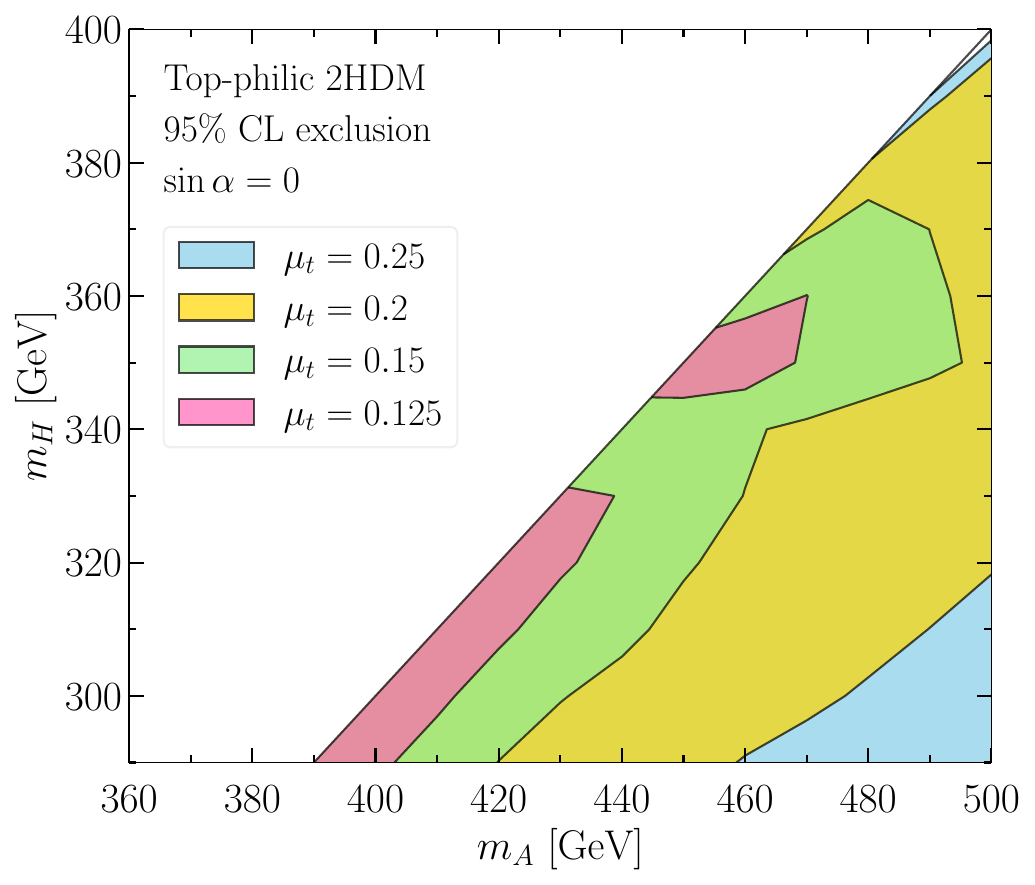}
\includegraphics[width=0.48\textwidth]{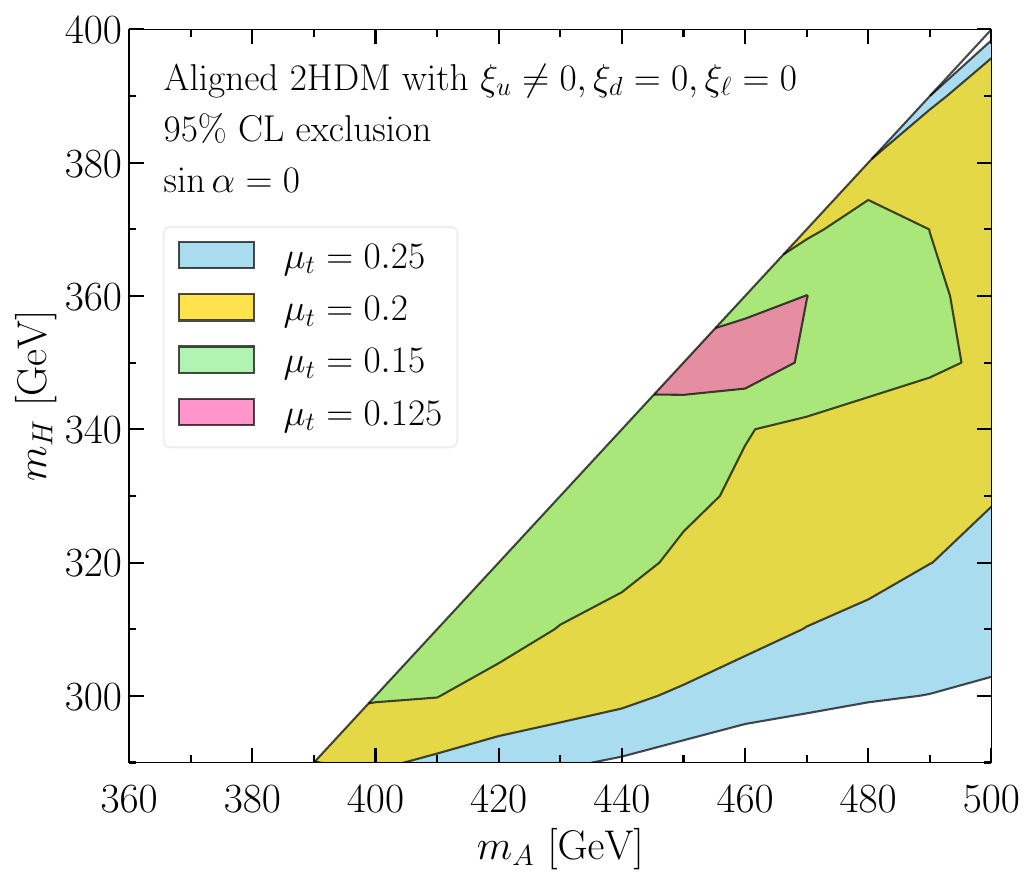}
\caption{Left: Regions in the $m_A$–$m_H$ plane for the top-philic 2HDM, assuming $\sin\alpha = 0$, that are excluded by our analysis for different values of the top Yukawa rescaling parameter $\mu_t$. Right: Corresponding exclusion in the aligned 2HDM realization with $\xi_u\neq 0$, i.e., both top and charm couplings. An extended version of this figure covering a larger $m_A$ range up to 800~GeV is provided in the Supplemental Material.}
\label{fig:THDM-top}
\end{figure*}

\subsection{Interpretation}
We now interpret these results in the top-philic realization of the two-Higgs-doublet model (2HDM). In 2HDMs, a second Higgs $SU(2)_L$ doublet, denoted $\Phi_1$~\cite{Branco:2011iw}, is added to the SM particle content. In the Higgs basis~\cite{Davidson:2005cw}, $\Phi_1$ does not acquire a VEV. The top-philic 2HDM constitutes a simplified realization within the general 2HDM with generic Yukawa couplings (Type~III)~\cite{Atwood:1996vj, Crivellin:2013wna}, obtained by assuming that $\Phi_1$ couples exclusively to the top quark. Normalizing the Yukawa interaction of the second doublet to the SM top Yukawa, the corresponding term reads $\mu_t Y_t \overline{Q}_3 \tilde\Phi_1 u_3$, where $\mu_t$ parametrizes the relative strength of the new interaction. For comparison, an approximately equivalent aligned 2HDM realization can be obtained by taking $\xi_u \neq 0$ while setting $\xi_d = \xi_\ell = 0$; we show results for both scenarios below.

In addition to the Higgs masses---$m_h=125$\,GeV, $m_H$ ($CP$-even), $m_A$ ($CP$-odd) and $m_{H^\pm}$ (charged)---the scalar sector is characterized by the mixing angle between the CP-even Higgs states ($\alpha$), and the quartic coupling $\lambda_5$ associated with the term $(\Phi_1^\dagger \Phi_2)^2$. The decoupling limit is achieved for $\alpha\to 0$ and $m_A,m_H,m_{H^\pm} \to \infty$. Assuming $m_A < m_{H^\pm} + m_W$, such that the decay $A \to H^\pm W$ is not possible, the branching ratio of $A \to ZH$ is controlled by $m_A$, $m_H$ and the couplings $\mu_t$ and $g_{AZH} \propto \cos\alpha$. The subsequent decay of $H$ is governed by its couplings to fermions and gauge bosons, $\mu_t$ and $g_{HVV} \propto \sin\alpha$. Thus, the $H \to t\bar{t}$ mode dominates close to the alignment limit $\sin\alpha \approx 0$.

We show in the left panel of Fig.~\ref{fig:THDM-top} the excluded regions in the $m_A$–$m_H$ plane for different values of the top Yukawa rescaling parameter $\mu_t$ in the top-philic 2HDM. In the right panel, corresponding to the aligned 2HDM with $\xi_u \neq 0$, the opening of the $H \to c\bar c$ channel leads to slightly weaker exclusions, while the qualitative features remain similar to the purely top-philic case. Most of this parameter space was previously not covered by dedicated $A \to ZH$ searches and remains consistent with perturbative unitarity and vacuum stability, Higgs signal strength measurements~\cite{Heo:2024cif}, electroweak precision data~\cite{ParticleDataGroup:2024cfk}, and bounds on the charged Higgs mass from inclusive weak radiative $B$-meson decays~\cite{Misiak:2017bgg}. Furthermore, for $\mu_t \lesssim 0.5$, the existing bounds from $A,H \to t\bar t$~\cite{ATLAS:2024vxm,CMS:2025dzq} and $tb$ production with $H^\pm \to tb$~\cite{CMS:2019rlz,ATLAS:2021upq} are generally satisfied.

For the best-fit region around $m_A \simeq 450$--460\,GeV and $m_H \simeq 290$\, GeV (see right plot in Fig.~\ref{fig:limit_combined}), the preference for a non-vanishing NP contribution can be accommodated for $0.16 \lesssim \mu_t \lesssim 0.33$ and $\sin\alpha \sim 0$; see Fig.~\ref{fig:THDM-top-best}. The $1\sigma$ and $2\sigma$ regions (shaded in blue) are shown together with the exclusions from the CMS and ATLAS searches for $A\to t\bar t$~\cite{ATLAS:2024vxm, CMS:2025dzq}. The corresponding best-fit line (purple) for the aligned realization is also displayed. 
The regions above the red and green lines are excluded by the ATLAS and CMS searches for $A \to t\bar{t}$~\cite{ATLAS:2024vxm,CMS:2025dzq}, respectively.

\section{Conclusions and Outlook}
In this article, we derived novel limits on the cross section of the process $pp \to A \to ZH$ with $H \to t\bar{t}$ in the low-mass region where one of the top quarks is off-shell. This part of parameter space was previously not explored by dedicated ATLAS and CMS analyses, despite being well motivated: a sizable mass splitting between the neutral Higgs states is more naturally realized near the EW scale than for higher masses, i.e., the consistency with perturbative unitarity is better (see Fig.~\ref{fig:perturbativity}).

\begin{figure}[htb!]
\centering
\includegraphics[width=0.47\textwidth]{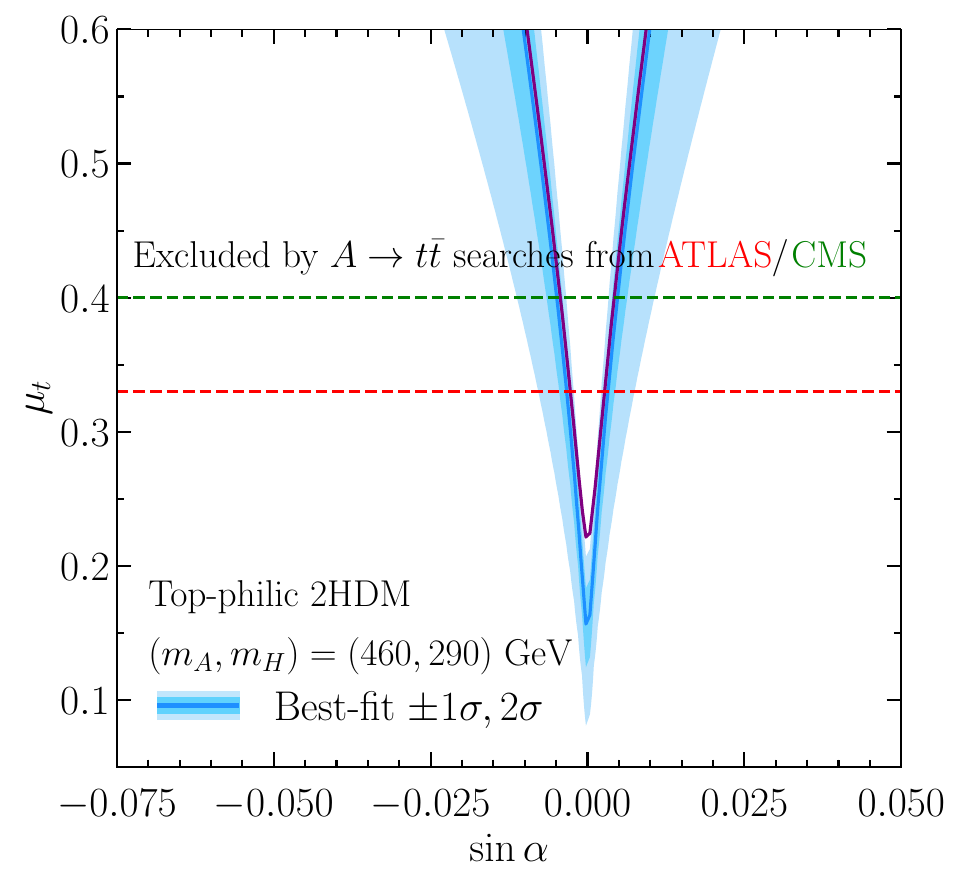}
\caption{Preferred $1\sigma$ and $2\sigma$ regions within the top-philic 2HDM for the best-fit point $(m_A, m_H) = (460, 290)~\mathrm{GeV}$, where a $2.5\sigma$ preference is observed. The purple curve indicates the best-fit line for the aligned 2HDM with $\xi_u \neq 0$, $\xi_d = \xi_\ell = 0$. The regions above the red and green lines are excluded by the ATLAS and CMS searches for $A \to t\bar{t}$~\cite{ATLAS:2024vxm,CMS:2025dzq}.}
\label{fig:THDM-top-best}
\end{figure}

We obtained limits on $\sigma(A \to ZH) \times \mathrm{Br}(H \to t\bar{t})$ in the range 0.12--0.62\,pb for $m_A > m_H + 100$\,GeV and $m_A$ between 360\,GeV and 500\,GeV, which are in good agreement with the dedicated ATLAS and CMS searches in the overlapping mass region. These strong bounds are obtained despite a preference (up to $\sim 2.5\sigma$) for an NP signal around $m_A \simeq 450$\,GeV--460\,GeV and $m_H \simeq 290$\,GeV.

Interpreting the results within the top-philic 2HDM, novel constraints are derived. Furthermore, the observed preference for a non-zero NP signal can be accommodated in the top-philic 2HDM for $0.16 \lesssim \mu_t \lesssim 0.33$. Future high-luminosity LHC measurements of differential $t\bar{t}Z$ production and dedicated searches for $A \to ZH$ with $H \to t\bar{t}$ will allow this region to be probed with greater precision, providing a direct test of the top-philic Higgs scenario like the aligned 2HDM with $\xi_{d,\ell}\approx0$ but $\xi_u\neq0$.
For completeness, we note that in 2HDMs with natural flavour conservation, the top Yukawa coupling is suppressed away from the low-$\tan\beta$ regime. As a consequence, in the mass region considered here, searches for $pp \to A \to ZH$ with $H \to b\bar b$~\cite{ATLAS:2020gxx} typically provide stronger constraints than the $H \to t\bar t$ channel. Our results are therefore complementary and mainly relevant for scenarios with enhanced top couplings.
In general, our analysis shows the importance and feasibility of searching for Higgs signals with off-shell top quarks in the final state.

\acknowledgments SA is supported by the Deutsche Forschungsgemeinschaft (DFG, German Research Foundation) under grant 396021762 -- TRR 257. AC is supported by a professorship grant of the Swiss National Science Foundation (Grant No.\ PP00P21\_76884). SPM and BM acknowledge the support of the Research Office of the University of the Witwatersrand. BM further acknowledges support from the South African Department of Science and Innovation through the SA-CERN program, and the National Research Foundation. We thank Thomas Biek\"otter for pointing out the ATLAS search for $A\to ZH \to \ell\ell bb$~\cite{ATLAS:2020gxx}, which excludes a sizable part of the type-I 2HDM parameter space. 

\bibliography{v0}

\clearpage
\newpage
\onecolumngrid

\setcounter{section}{0}
\renewcommand{\thesection}{S\arabic{section}}

\begin{center}
{\large\bf Supplemental Material}
\end{center}

\section{CMS and ATLAS Event Selections}
\label{app:event}
We now summarize the event selections applied in our recast to reproduce the fiducial $t\bar t Z$ regions used by CMS and ATLAS for their differential measurements.

To reproduce the CMS $t\bar t Z$ signal regions \cite{CMS:2024mke}, we require events with at least three charged leptons ($e,\mu$). Leptons are required to satisfy $|\eta|<2.5$ for electrons and $|\eta|<2.4$ for muons, with transverse-momentum thresholds of $p_T>25$, $15$, and $10$~GeV for the three leading leptons. A $Z$-boson candidate is reconstructed from a same-flavor opposite-sign lepton pair with invariant mass $70<m_{\ell\ell}<110$~GeV, while the third lepton is interpreted as originating from the $W$ boson decay. Jets are reconstructed using the anti-$k_T$ algorithm with $R=0.4$, requiring $p_T^j>25$~GeV and $|\eta_j|<5.0$, and are required to be separated from leptons by $\Delta R(j,\ell)>0.4$. Events are required to contain at least two jets, of which at least one must be $b$-tagged.

To reproduce the ATLAS $t\bar t Z$ fiducial selections used for the differential measurements, we follow the particle-level definitions given in Ref.~\cite{ATLAS:2023eld}. In the trilepton channel, we require exactly three charged leptons ($e,\mu$) with $p_T>27,20,15$~GeV and total charge $\pm1$, at least one opposite-sign same-flavor (OSSF) lepton pair with $|m_{\ell\ell}-m_Z|<10$~GeV, and $m_{\mathrm{OSSF}}>10$~GeV for all OSSF combinations. We further require at least three jets with $p_T^j>25$~GeV and $|\eta_j|<2.5$, and at least one $b$-tagged jet. In the tetralepton channel, we require exactly four leptons with $p_T>27,7,7,7$~GeV and total charge zero, at least one OSSF lepton pair with $|m_{\ell\ell}-m_Z|<20$~GeV, and $m_{\mathrm{OSSF}}>10$~GeV for all OSSF combinations, together with at least two jets with $p_T^j>25$~GeV and $|\eta_j|<2.5$, and at least one $b$-tagged jet.

\begin{figure*}[h]
\centering
\includegraphics[width=0.95\textwidth]{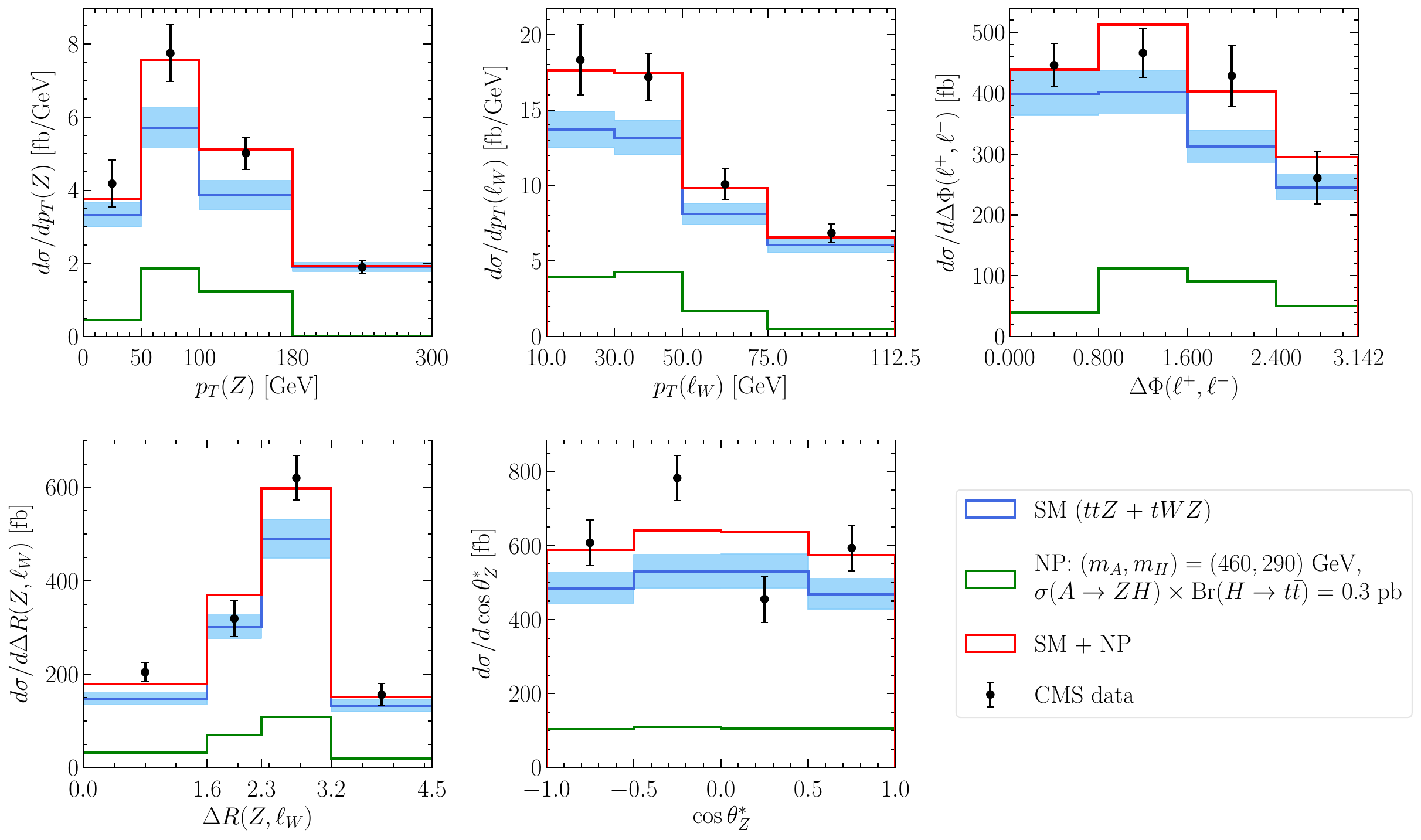}
\caption{$t\bar tZ+tWZ$ differential cross sections for various observables measured by CMS. The error bars indicate the total experimental error, while the shaded blue area corresponds to the theory uncertainty on the SM prediction as provided by the CMS differential measurement, including contributions from scale variations, PDFs, and parton-shower modelling. The NP contributions are shown for $(m_A,m_H)=(460, 290)$\,GeV and $\sigma(A \to ZH) \times {\rm Br}(H \to t\bar t) = 0.3$\,pb, corresponding to the combined (ATLAS and CMS) best-fit point. The corresponding couplings in the top-philic 2HDM depend on $\alpha$ and $\mu_t$, with the region associated with this benchmark shown in Fig.~5 of the main text.}
\label{fig:CMS_dist}
\end{figure*}

\section{Differential Cross Sections for CMS and ATLAS Observables}
\label{app:dist}
The differential cross sections for the observables measured by CMS and ATLAS are shown in Fig.~\ref{fig:CMS_dist} and Fig.~\ref{fig:ATLAS_dist}, respectively. The data and the SM predictions are taken from each analysis, while the NP predictions correspond to $(m_A,m_H)=(460, 290)$\,GeV for the (combined) best-fit point $\sigma(A \to ZH) \times {\rm Br}(H \to t\bar t) \approx 0.3$\,pb.

\begin{figure*}[htb!]
\centering
\includegraphics[width=0.95\textwidth]{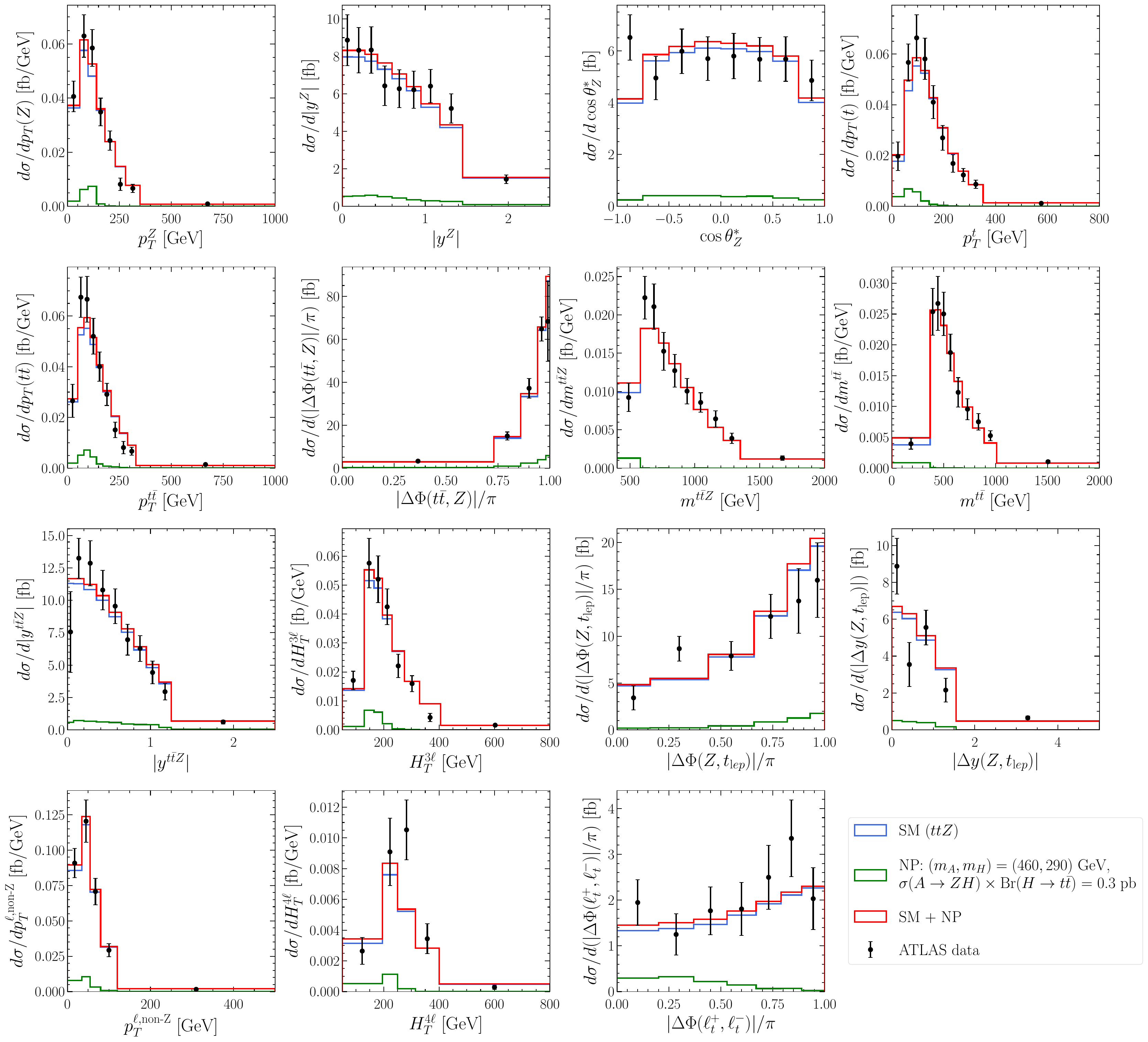}
\caption{Differential $t\bar tZ$ cross sections unfolded to particle level for different observables measured by ATLAS. The error bars indicate the total experimental uncertainties. The NP contributions are shown for $(m_A,m_H)=(460, 290)$\,GeV and $\sigma(A \to ZH) \times {\rm Br}(H \to t\bar t) = 0.3$\,pb, corresponding to the combined (ATLAS and CMS) best-fit point.}
\label{fig:ATLAS_dist}
\end{figure*}


\section{Extended limits in the $m_A$--$m_H$ plane}
\label{app:extended}
Finally, Fig.~\ref{fig:extended} displays the extended $m_A$--$m_H$ plane with the resulting 95\%~CL upper limits on $\sigma(A \to ZH) \times {\rm Br}(H \to t\bar{t})$, extending the results of Fig.~3 of the main text. In addition, Fig.~\ref{fig:THDM-top-extended} displays the corresponding extension of the 2HDM interpretation, illustrating the largest excluded pseudoscalar masses for large values of $\mu_t$.

\begin{figure*}[htb!]
\centering
\includegraphics[width=0.95\textwidth]{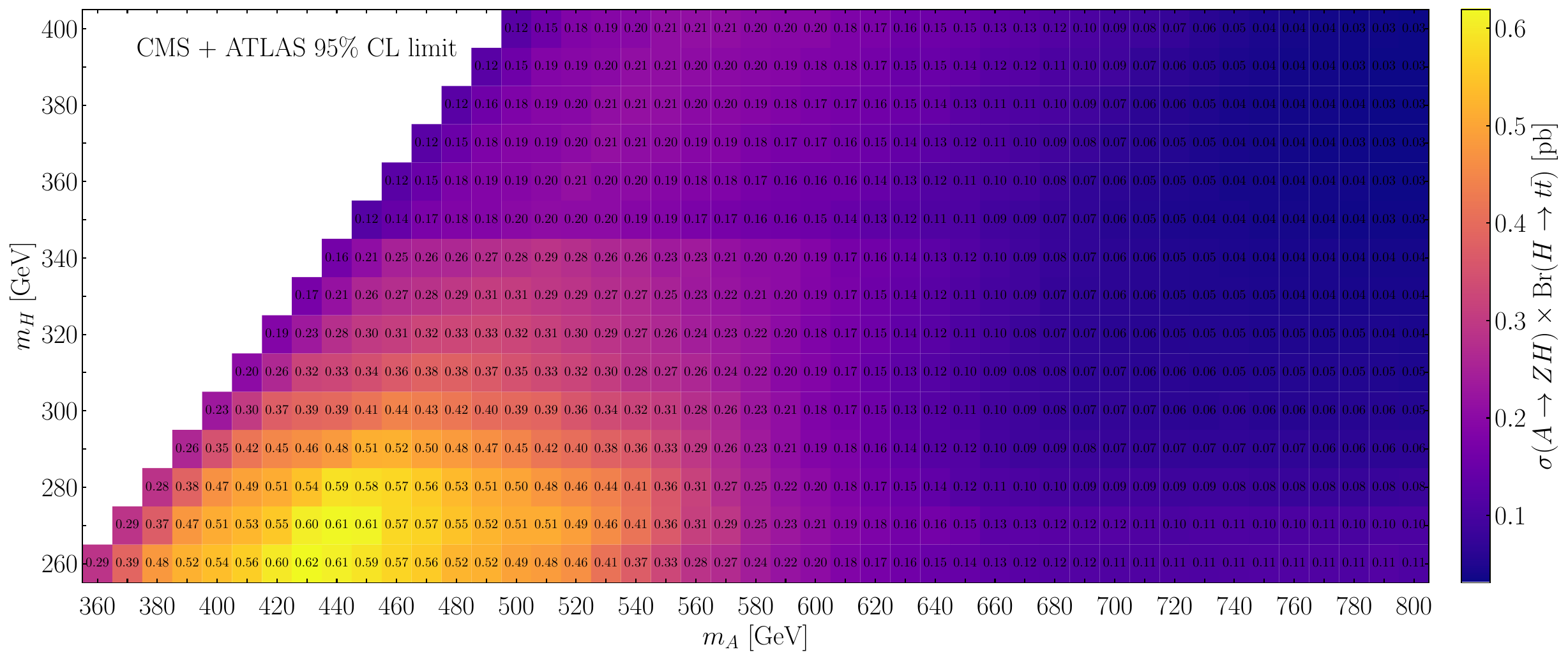}
\caption{Extended 95\%~CL upper limits on $\sigma(A \to ZH) \times \mathrm{Br}(H \to t\bar{t})$ in units of pb in the $m_A$--$m_H$ plane with $m_A - m_H \geq 100~\mathrm{GeV}$. This plot extends the results shown in Fig.~3 of the main text to a wider mass range.}
\label{fig:extended}
\end{figure*}

\begin{figure*}[t!]
\centering
\includegraphics[width=0.41\textwidth]{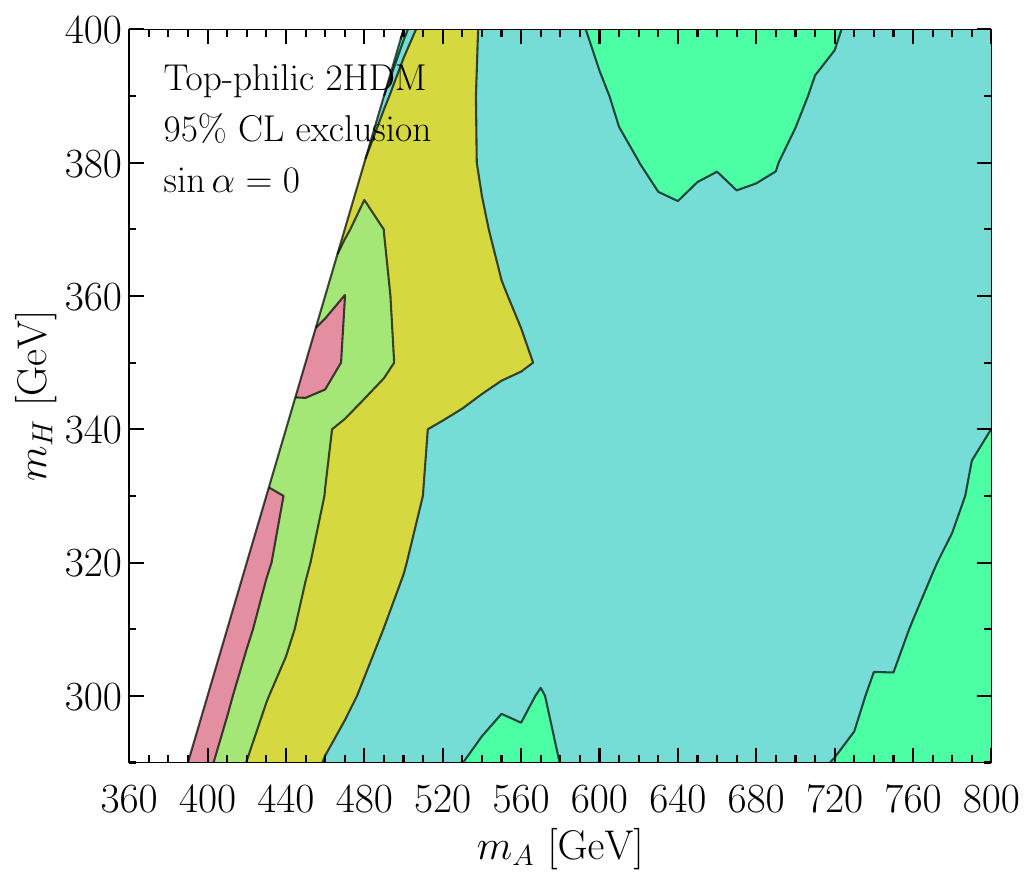}
\includegraphics[width=0.5\textwidth]{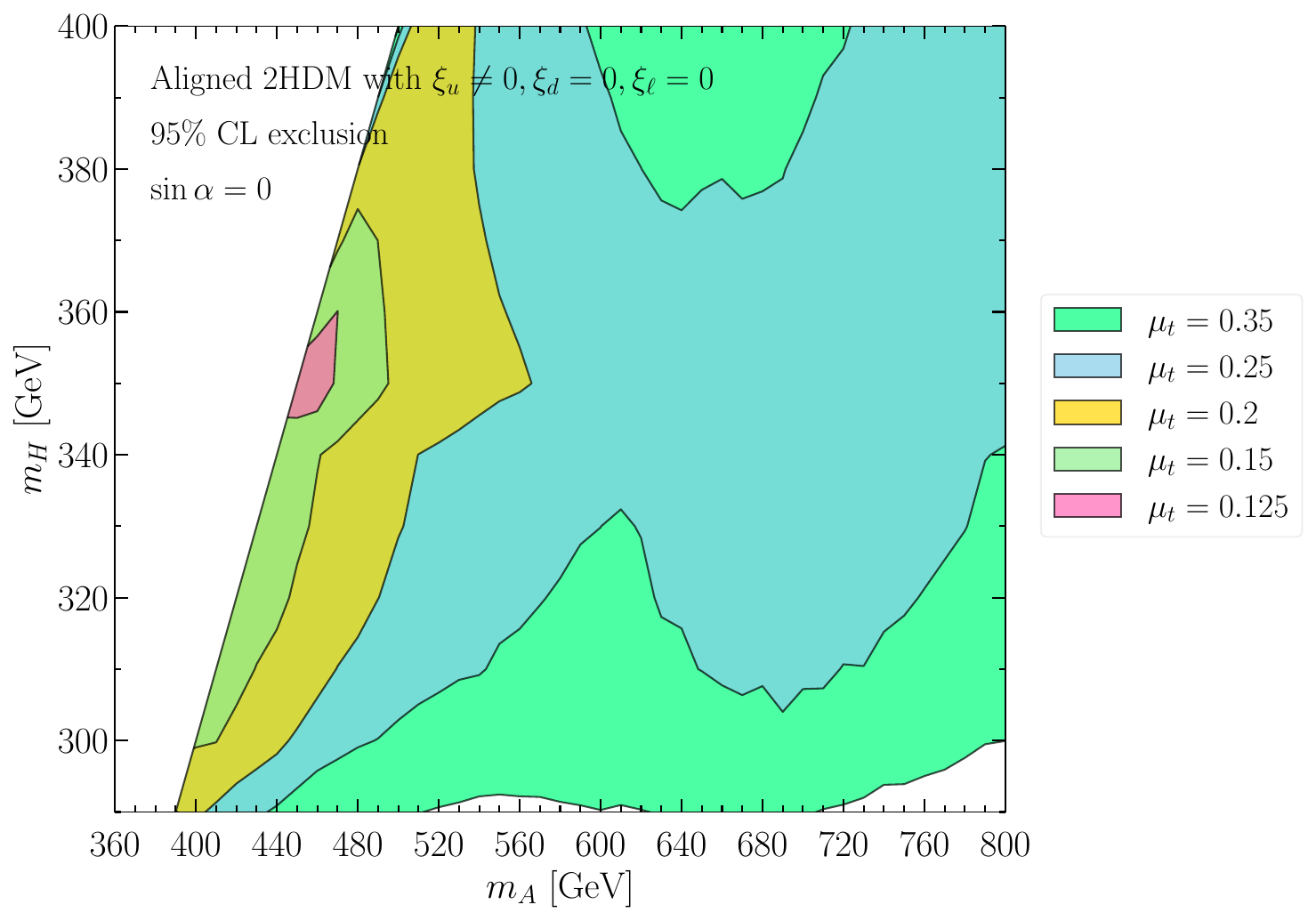}
\caption{Same as Fig.~4, but extended to $m_A<800$~GeV to illustrate the largest excluded masses for large $\mu_t$.}
\label{fig:THDM-top-extended}
\end{figure*}

\end{document}